\documentclass[conference]{IEEEtran}
\usepackage{amssymb}
\usepackage[algoruled,vlined,linesnumbered]{algorithm2e}
\usepackage{comment}
\usepackage{amsthm}

\theoremstyle{definition}

\theoremstyle{remark}

\usepackage{diagbox}
\usepackage{bm}
\usepackage{cite} 
\usepackage{amsmath}
\usepackage{float}
\usepackage{graphicx}
\usepackage{subfigure}
\usepackage{xcolor}
\usepackage{array}
\usepackage{hyperref}
\title{Meta-learning Based  Beamforming Design for MISO Downlink}
\author{\IEEEauthorblockN{Jingyuan Xia}
\IEEEauthorblockA{School of Electrical Technology\\
National University of Defense Technology\\
Changsha, China 410072\\
Email: j.xia16@imperial.ac.uk}
\and
\IEEEauthorblockN{Deniz Gunduz}
\IEEEauthorblockA{Department of Electrical and Electronic Engineering
\\Imperial College London\\
London, UK\\
Email: d.gunduz@imperial.ac.uk}
}

\begin{document}
\maketitle

\begin{abstract}
Downlink beamforming is an essential technology for wireless cellular networks; however, the design of beamforming vectors that maximize the weighted sum rate (WSR) is an NP-hard problem and iterative algorithms are typically applied to solve it. The weighted minimum mean square error (WMMSE) algorithm is the most widely used one, which iteratively minimizes the WSR and converges to a local optimal. Motivated by the recent developments in meta-learning techniques to solve non-convex optimization problems, we propose a meta-learning based iterative algorithm for WSR maximization in a MISO downlink channel. A long-short-term-memory (LSTM) network based meta-learning model is built to learn a dynamic optimization strategy to update the variables iteratively. The learned strategy aims to optimize each variable in a less greedy manner compared to WMMSE, which  updates variables by computing their first order stationary points at each iteration step. The proposed algorithm outperforms WMMSE significantly in the high signal to noise ratio (SNR) regime and achieves comparable performance when the SNR is low. Our code is available at \href{https://github.com/XiaGroup/MLBF}{https://github.com/XiaGroup/MLBF}.
\end{abstract}

\section{Introduction}
Downlink beamforming is an essential technology for multi-antenna cellular networks. Optimization of the beamforming vectors is typically formulated as a weighted sum rate (WSR) maximization problem under a total transmit power constraint. However, the WSR maximization problem is non-convex and NP-hard \cite{luo2008dynamic}. Therefore, it is typically solved by iterative methods which either adopt convex approximations \cite{peel2005vector,ng2010linear,kibria2013coordinated}, or convert the original problem into an alternative formulation with closed-form solutions \cite{christensen2008weighted,schmidt2009minimum,shi2011iteratively}. Among them, the iterative weighted minimum mean square error (WMMSE) algorithm is the most widely used approach that provides superior performance in most scenarios.  

The WMMSE algorithm converts the WSR maximization problem into a weighted sum mean square error minimization problem under the total transmit power constraint. The equivalent minimization problem is convex with respect to each individual variable when the others are fixed. Therefore, a locally optimal solution can be obtained by minimizing the objective function with respect to individual variables in an iterative manner.  The WMMSE algorithm has been shown to achieve state-of-the-art performance in a wide range of scenarios, and is often considered as the benchmark in the literature.

Our goal in this paper is to obtain a less greedy and more flexible iterative updating strategy, such that the solution of each individual variable moves away from the locally optimal solution and  closer to the globally optimal solution. To achieve such an adaptive algorithm, we incorporate machine learning techniques.     
Using machine-learning-based solutions to improve the WMMSE algorithm has received significant recent attention. However, the common approaches are built on an end-to-end learning model, where the wireless channel coefficients are given as input to a neural network, which then outputs the estimation of the transmitter beamformer  \cite{sun2018learning,xia2019deep,huang2018unsupervised}. These approaches convert the original model-based problem into a data-driven learning-based problem, and the performance depends on the the network architecture.  Alternatively, more recent works in \cite{Pellaco2020Deepunfolding,Chowdhury2020UnfoldingWMMSE} propose  unfolding the WMMSE algorithm in order to design a model-inspired computational structure for the neural network, and learn a specific {optimization rule for updating the WMMSE parameters by training on a set of channel samples.} The unfolded WMMSE algorithm follows the original mathematical model of the WSR optimization problem, and designs a specific network structure to map the variables and the parameters in the original problem; therefore, the optimization of the WSR problem is converted into the optimization of the designed neural network. However, the performance of the aforementioned approaches \cite{yuan2020transfer,sun2018learning,xia2019deep,Pellaco2020Deepunfolding,Chowdhury2020UnfoldingWMMSE} do not surpass the WMMSE algorithm.

An alternative approach is to utilize meta-learning techniques to solve the underlying optimization problem. The work in \cite{andrychowicz2016learning} uses meta-learner networks to learn the variable updating strategy in the form of gradient descent, which surpasses the benchmark approaches including Adam \cite{kingma2014adam} and RMSProp \cite{hinton2012neural} in convergence speed. In \cite{li2016learning,li2017learning}, meta-learning is applied to the design of solution algorithms for different gradient-descent-based optimization problems. Another popular model-agnostic meta-learning (MAML) algorithm \cite{finn2017meta,finn2017model} uses meta-learning for finding a superior initialization, which can lead to a faster convergence. This idea has been recently employed in \cite{yuan2020transfer} for the WSR maximization problem, where there can be a mismatch between test and training SNRs. Meta-learning allows fine-tuning the NN parameters with few training samples. A meta-learning based global scope optimization (GSO) method is proposed in \cite{xia2020meta} for solving non-convex multi-variable optimization problems. The GSO algorithm exhibits significantly better performance in solving non-convex problems, such as matrix completion without rank information or Gaussian mixture model with high dimensionality, than the traditional alternating minimization-based methods.

\begin{figure}[tbp]
    \centering
    \includegraphics[width=8.5cm]{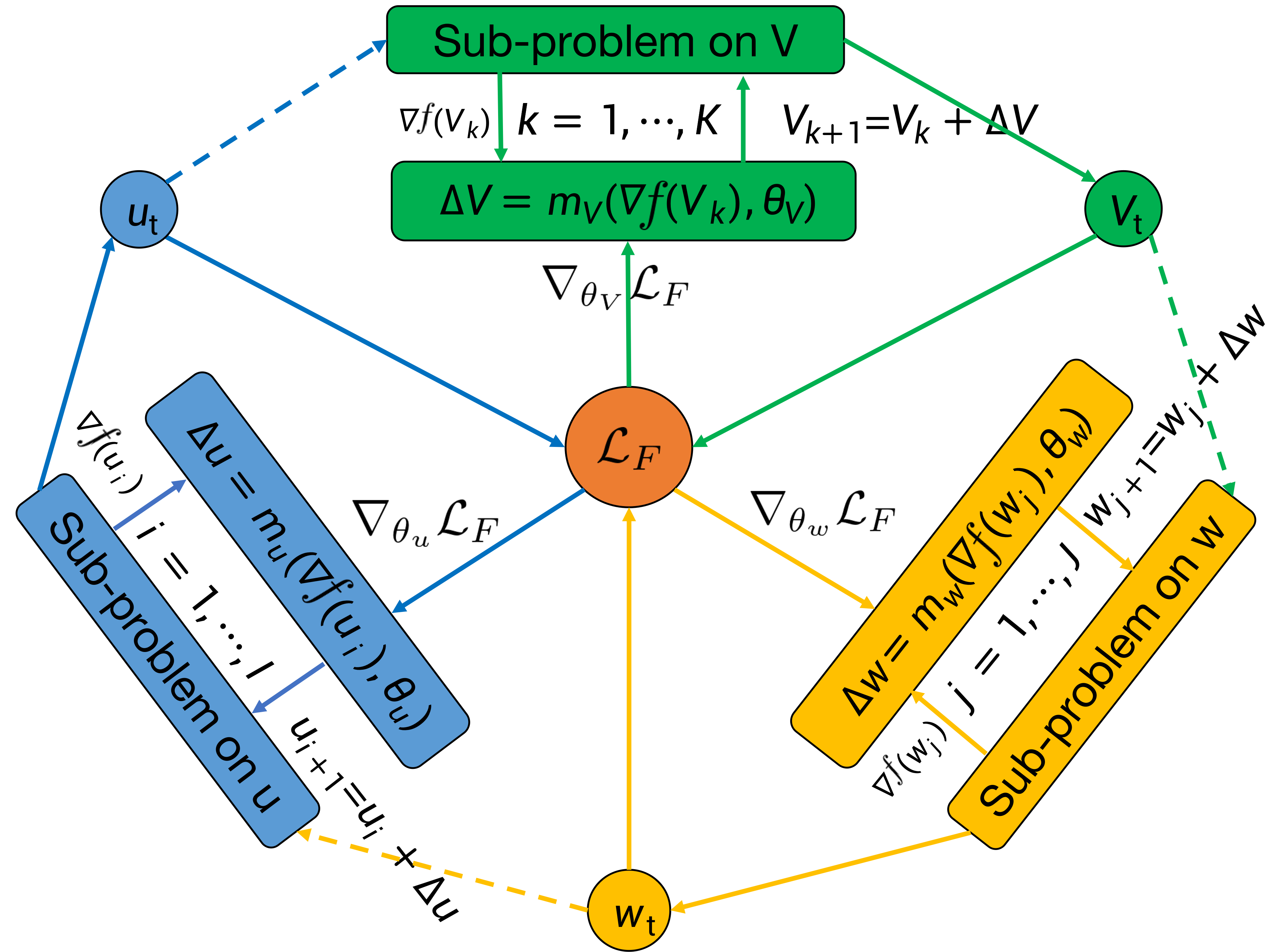}
    \caption{The MLBF algorithm builds a bridge of information flow between the updates on each variable and the global objective function (denoted by $\mathcal{L}_F$). Each color represents the gradient flow for one individual variable.
    We use solid arrows to denote information flow, and dashed arrows to indicate the lack of gradient flow. It shows that the MLBF algorithm achieves a circular information flow between the variable update steps within each sub-problem and the global objective function throughout iterations.}
    \label{fig:GSO_structure}
\end{figure}

In this paper, we propose a meta-learning aided beamformer (MLBF) algorithm to solve the WSR maximization problem. The MLBF algorithm is built upon the iterative  block-coordinate descent method, but instead of greedily optimizing each individual variables at each step, it can adaptively optimize each variable with respect to the geometry of the objective function. The proposed MLBF algorithm is established by three long-short-term-memory (LSTM) neural networks corresponding to the three complex variables to be optimized. 
The general structure of the MLBF algorithm for the WSR maximization problem is depicted in Fig.\ref{fig:GSO_structure}. The notations and the detailed explanation will be presented in Section \ref{sec:GSO}. It can be seen that the MLBF algorithm builds a bridge between the accumulated global loss function $\mathcal{L}_F$ and the update rules on each variable. This allows the gradient flow to circulate between the inner update loops for individual variable (referring to the update steps $i$, $j$ and $k$ for each variable) and the outer iterative steps for different variables (referring to different $t$).

The main contribution of this work is the proposed MLBF algorithm that can learn a less greedy update strategy for the variables, and therefore, achieves a better performance than the WMMSE algorithm.  The extensive numerical results show that the proposed  MLBF algorithm outperforms the WMMSE algorithm in terms of WSR when the SNR is high and performs equally at low SNR. We observe that the gain from the proposed MLBF algorithm in terms of the achieved WSR compared to the WMMSE algorithm increase with the channel SNR. 

\section{Problem Formulation}
In this paper, we study a multiple-input single output (MISO) downlink channel, where a base station with $M$ transmit antennas serves $N$ single-antenna users.  Let  $x_{i}\sim\mathcal{CN}(0,1)$ denote the independent data symbol transmitted to the $i^{\text{th}}$ user, while $\bm{h}_i\sim\mathcal{CN}(\bm{0},\bm{I}_{M})$ denote the channel between the base station and the $i^{\text{th}}$ user. Then the received signal at the $i^{\text{th}}$ user is given by \begin{equation}
    y_{i}=\bm{h}_{i}^{H}\bm{v}_{i}x_{i}+\sum_{j=1,j\not=i}^{N}\bm{h}_{i}^{H}\bm{v}_{j}x_{j}+n_{i},
\end{equation}
where $\bm{v}_i\in \mathbb{C}^{M}$ is the transmit beamforming vector for user $i$ and $n_{i}\in\mathcal{CN}(0,\sigma^2)$ denotes a complex circularly symmetric zero mean Gaussian noise with variance $\sigma^2$. The signal-to-interference-plus-noise-ratio (SINR) is given by\begin{equation}\label{SINR}
    \text{SINR}_i=\frac{|\bm{h}_i^H\bm{v}_i|^2}{\sum_{j=1,j\not=i}|\bm{h}_i^H\bm{v}_j|^2+\sigma^2}.
\end{equation}
Given the channel knowledge, the WSR maximization problem of the downlink channel with respect to a total transmit power constraint is formulated as \begin{equation}\label{WSR}
\begin{aligned}
       \max_{\bm{V}} \sum_{i=1}^N \alpha_i\log_2(1+\text{SINR}_i)\\
    \text{s.t.}\; \text{Tr}(\bm{V}\bm{V}^H)\leq P, 
\end{aligned}
\end{equation}where $\alpha_i$ is the user weight specified by the system designer, $P$ is the maximum transmit power, $\bm{V}\triangleq[\bm{v}_1,\ldots,\bm{v}_N]^{T}$, and $\text{Tr}(\cdot)$ denotes the trace operator.

\section{WMMSE Algorithm}
As mentioned earlier, the WSR maximization problem in (\ref{WSR}) is NP hard and non-convex. The iterative WMMSE algorithm is typically implemented to solve it in the following steps: First, problem (\ref{WSR}) is converted into the equivalent  weighted sum mean square error minimization problem:
\begin{equation}\label{minWSR}
\begin{aligned}
   \min_{\bm{u},\bm{w},\bm{V}}\sum_{i=1}^{N}\alpha_i(w_{i}e_{i}-\log_{2}w_i)\\
   \text{s.t.}\; \text{Tr}(\bm{V}\bm{V}^H)\leq P,
\end{aligned}
\end{equation}where $e_{i}$ is the mean-square error given by 
\begin{equation}\label{e_i complex}
\begin{aligned}
    \\e_{i}&=\mathbb{E}_{\bm{x},n_i,}[|\hat{x_{i}}-x_i|^2]
    \\&=|u_{i}\bm{h}_{i}^H\bm{v}_i-1|^2+\left(\sum_{j\not=i,j=1}^N|u_{i}\bm{h}_{i}^H\bm{v}_j|^2\right)+\sigma^2|u_i|^2,
\end{aligned}
\end{equation}
$u_i$ denotes the receiver gain, $w_i$ is a positive user weight, $\bm{w}=[w_1,\ldots,w_N]^T$, $\bm{u}=[u_1,\ldots,u_N]^T$, and {$n_i$ is the noise term}. Although this new formulation is also non-convex, problem (\ref{minWSR}) is convex with respect to each individual variable when the others are fixed. Therefore, this new formulation can be solved by block coordinate descent. Indeed, a closed-form solution with respect to the first-order stationary point for each variable can be obtained. Then, the WMMSE algorithm iteratively updates each variable $i=1,\ldots,N$ by computing their first order stationary points as follows
\begin{equation}\label{w_update}
 w_i=\frac{\sum_{j=1}^N|\bm{h}_{i}^H\bm{v}_j|^2+\sigma^2}{\sum_{j\not=i,j=1}^N|\bm{h}_{i}^H\bm{v}_j|^2+\sigma^2},   
\end{equation}

\begin{equation}\label{u_update}
u_i=\frac{\bm{h}_{i}^H\bm{v}_i}{\sum_{j=1}^N|\bm{h}_{i}^H\bm{v}_j|^2+\sigma^2},   
\end{equation}

\begin{equation}\label{V_update}
 \bm{v}_i=\alpha_i u_i w_i \bm{h}_i(\bm{A}+\mu\bm{I})^{-1},   
\end{equation}where $\bm{A}\triangleq\sum_{i=1}^N\alpha_i w_i |u_i|^2\bm{h}_i\bm{h}_i^H$, and $\mu\geq0$ is a Langrange multiplier. 
{In a nutshell, the WMMSE algorithm first
initializes $\bm{V}$ so that it satisfies the power constraint, then iteratively update $\bm{w}$, $\bm{u}$, and $\bm{V}$ following the equations (\ref{w_update})-(\ref{V_update}) until a stop criterion is met. Parameter $\mu$ is computed by bisection search.} Further details of the WMMSE algorithm can be found in \cite{shi2011iteratively}.   

\section{A Meta-learning based Beamforming (MLBF) Algorithm}\label{sec:GSO}
We propose a meta-learning based WMMSE algorithm to learn an adaptive updating strategy of each variable in problem (\ref{minWSR}). In particular, for each variable, the update term is generated by a meta-learner neural network whose parameters are continuously  updated with the objective of minimizing the global loss function $F(\bm{u},\bm{w},\bm{V})$. We refer to the minimization of $F(\bm{u},\bm{w},\bm{V})$ as the overall problem while to the minimization of  $f(\bm{u})$,$f(\bm{w})$ and $f(\bm{V})$ as the three sub-problems to be optimized sequentially in each iteration. Following (\ref{minWSR}) the overall problem can be written as \begin{equation}\label{global loss function}
\min F(\bm{u},\bm{w},\bm{V})\triangleq\sum_{i=1}^{N}\alpha_i(w_{i}e_{i}-\log_{2}w_i)+\mu \text{Tr}(\bm{V}\bm{V}^{H})-\mu P.
\end{equation} 
Three LSTM networks $m_{\bm{V}}$, $m_{\bm{w}}$ and $m_{\bm{u}}$ are established for updating  $\bm{V}$, $\bm{w}$, and $\bm{u}$, respectively. We define the variable update process within each sub-problem as inner loops and the iterative steps over different sub-problems as outer loops. We correspondingly denote by $\bm{\theta}_{\bm{V}}$, $\bm{\theta}_{\bm{u}}$ and $\bm{\theta}_{\bm{w}}$ the parameters of the three LSTM networks, and by $\bm{C}_{\bm{V}}$, $\bm{C}_{\bm{u}}$,  and $\bm{C}_{\bm{w}}$  their cell states. 
The inputs of the LSTM networks are the gradients of the sub-problems and the previous states of the variables which are represented by the corresponding $\nabla f()$ and $\bm{C}$, respectively. The outputs of the LSTM networks include updating steps for variables, along with the update of the cell state. The cell states contain the previous variables' information, and can be adaptively adjusted by the inner control gates inside the corresponding LSTM networks. Further details can be found in \cite{xia2020meta}.
Denoting the inner loop update steps by $k=1,\ldots,K$, $i=1,\ldots,I$, and $j=1,\ldots,J$ as  superscripts and the outer loop steps by $t=1,\ldots,T$ as subscripts for each sub-problem, where $K$, $I$ and $J$ denote the maximum number of steps of the inner loops for each variable, and $T$ denotes the maximum number of steps of the outer loop, respectively. The meta-learner networks $m_{\bm{V}}$, $m_{\bm{w}}$ and $m_{\bm{u}}$ carry out variable updates in the inner loops in the following form:
\begin{equation}
    \begin{aligned}
    \bm{V}^{(k)}=\bm{V}^{(k-1)}+m_{\bm{V}}(\nabla f(\bm{V}^{(k-1)}),\bm{C}_{\bm{V}^{(k-1)}},\bm{\theta}_{\bm{V}}),\\
    \bm{u}^{(i)}=\bm{u}^{(i-1)}+m_{\bm{u}}(\nabla f(\bm{u}^{(i-1)}),\bm{C}_{\bm{u}^{(i-1)}},\bm{\theta}_{\bm{u}}), \\
    \bm{w}^{(j)}=\bm{w}^{(j-1)}+m_{\bm{w}}(\nabla f(\bm{w}^{(j-1)}),\bm{C}_{\bm{w}^{(j-1)}},\bm{\theta}_{\bm{w}}). 
    \end{aligned}
\end{equation} Within the inner loops, the parameters of these three networks are fixed, and are used to generate the updates  for the variables as the input of the variable update function. In the outer loops, we update the network parameters through backpropagation with respect to the accumulated global loss, given by\begin{equation}
\mathcal{L}_{F}^s=\frac{1}{t_{up}}\sum_{t_s=(s-1)t_{up}+1}^{st_{up}}\omega_{t_s}F(\bm{u}_{t_s},\bm{w}_{t_s},\bm{V}_{t_s}),\label{eq:Mean Global Loss}
\end{equation} 
where $t_{up}$ is the update interval; that is, for every $t_{up}$ outer loop iterations, the accumulated global loss  $\mathcal{L}_{F}^s$ is used to update $\bm{\theta}_{\bm{V}}$, $\bm{\theta}_{\bm{u}}$ and $\bm{\theta}_{\bm{w}}$, $s=1,2,\ldots,S$, and $\omega_{t_s}\in\mathbb{R}_{\geq0}$ denotes the weights associated with each outer loop step. $T=St_{up}$ denotes the maximum number of outer loop steps. Every $t_{up}$ outer loop iterations, the parameters of the LSTM networks are updated by the Adam \cite{kingma2014adam} optimizer using the accumulated global loss $\mathcal{L}_{F}^s$:
\begin{equation}
    \begin{aligned}
        \bm{\theta}_{\bm{V}}^{s+1}=\bm{\theta}_{\bm{V}}^s+\alpha_{\bm{V}}\cdot\mathrm{Adam}(\bm{\theta}_{\bm{V}}^s, \nabla_{\bm{\theta}_{\bm{V}}^s}\mathcal{L}_{F}^s),\\
        \bm{\theta}_{\bm{u}}^{s+1}=\bm{\theta}_{\bm{u}}^s+\alpha_{\bm{u}}\cdot\mathrm{Adam}(\bm{\theta}_{\bm{u}}^s, \nabla_{\bm{\theta}_{\bm{u}}^s}\mathcal{L}_{F}^s), \\
        \bm{\theta}_{\bm{w}}^{s+1}=\bm{\theta}_{\bm{w}}^s+\alpha_{\bm{w}}\cdot\mathrm{Adam}(\bm{\theta}_{\bm{w}}^s, \nabla_{\bm{\theta}_{\bm{w}}^s}\mathcal{L}_{F}^s),   
    \end{aligned}
\label{update theta}
\end{equation}
where $\alpha_{\bm{V}}$, $\alpha_{\bm{u}}$ and $\alpha_{\bm{w}}$ denote the learning rate for the LSTM networks $m_{\bm{V}}$, $m_{\bm{u}}$ and $m_{\bm{w}}$, respectively. 

To strictly satisfy the total power constraint when neural networks are used to update the variables,  we  define $\mathcal{D}\triangleq\{\bm{V}|\text{Tr}(\bm{V}\bm{V}^{H})\leq P\}$ as the set of beamforming matrices that satisfy the power constraint. Then, we apply the following projection  of the beamforming matrix $\bm{V}$ to set $\mathcal{D}$ in each outer loop step when the inner loop updating for $\bm{V}$ is completed.
\begin{equation}
    \Omega_{\mathcal{D}}\{\bm{V}\}=\left\{\begin{array}{l}\bm{V},\;\;\;\;\;\;\;\;\;\;\; \text{if}\;\bm{V}\in\mathcal{D}, \\
\frac{\bm{V}}{||\bm{V}||_F}\sqrt{P}, \text{otherwise}. 
\end{array}\right.
\end{equation} 
In conclusion, the MLBF algorithm  builds a connection between the variable update functions and the global loss function. Specifically, at the outer loops, the three networks update their parameters with the objective of maximizing the WSR (equivalently minimizing the global loss function in (\ref{global loss function})), which
then determine the generation of the variable update terms with the 
inner loops. In this way, the knowledge of the global loss is  reflected on the update rules of each variable within the inner loops; that is,  $\bm{\theta}_{\bm{V}}$, $\bm{\theta}_{\bm{u}}$ and $\bm{\theta}_{\bm{w}}$ convey the global loss knowledge from the outer loops for the variable update functions. This allows the individual variable update functions to be  less-greedy, and to return updates that are more aligned with the global loss function instead of the first-order stationary points of the corresponding sub-problems. Therefore, the learned updating strategy may not necessarily minimize the loss function of each sub-problem, but the overall problem could be better accommodated compared to the WMMSE algorithm. 
The general structure of the proposed MLBF algorithm for the WSR maximization problem is presented in \textbf{Algorithm} \ref{alg:meta-WMMSE}. 

We highlight that, here the proposed MLBF algorithm is applied directly as a solution algorithm. For each WSR problem to be optimized, the LSTM networks' parameters of the implemented MLBF algorithm start from scratch and are trained during the iterative steps, such that the finally obtained variables are assigned as the solutions. We note that the computational complexity of the MLBF algorithm is higher than the WMMSE algorithm, nevertheless, it is directly used as a solution algorithm for WSR problem. In this way, the proposed MLBF approach is less time-consuming and resource-dependent to be applied than the previous deep-learning based approaches \cite{Pellaco2020Deepunfolding,xia2019deep,huang2018unsupervised,Chowdhury2020UnfoldingWMMSE}, as the training process is essentially embedded in the solution process while the traditional deep-learning based approaches have to train the parameters of the networks with a set of training samples for a certain period of time (which is typically quite long)  before they can be used to solve a particular instance of the problem.    

\begin{algorithm}[htbp]
    \SetAlgoLined
    \textbf{Given:} global loss function $F(\bm{u},\bm{w},\bm{V})$, $\bm{H}$.
    
    \textbf{Initialized:} $\bm{V}_0$, $\bm{u}_0$, and $\bm{w}_0$.
    
    \For{$t\gets 1$,2, $\ldots$, T}{
    $i, j, k, s = 1$

    \While{$i\leq I$}
        {$\Delta \bm{u}=m_{\bm{u}}(\nabla f(\bm{u}^{(i-1)}),\bm{C}_{\bm{u}}^{(i-1)},\bm{\theta}_{\bm{u}}^s)$
        
        $\bm{u}^{(i)}\leftarrow \bm{u}^{(i-1)}+\Delta \bm{u}$\
        $i = i + 1$
        }
    $\bm{u}_t = \bm{u}^{(I)} $ 
    
    generate $f(\bm{w}) \text{with}\; \bm{u}_t,\bm{V}_{t-1}$

    \While{$j \leq J$}
        {$\Delta \bm{w} =m_{\bm{w}}(\nabla f(\bm{w}^{(j-1)}),\bm{C}_{\bm{w}}^{(j-1)},\bm{\theta}_{\bm{w}}^s)$
        
        $\bm{w}^{(j)}\leftarrow \bm{w}^{(j-1)}+\Delta \bm{w}$\
        $j = j + 1$
    }

    $\bm{w}_t = \bm{w}^{(J)} $ 
    
    generate $f(\bm{V}) \text{with}\; \bm{w}_t, \bm{u}_t$
    
       \While{$k \leq K$}
        {$\Delta \bm{V} =m_{\bm{V}}(\nabla f(\bm{V}^{(k-1)}),\bm{C}_{\bm{V}}^{(k-1)},\bm{\theta}_{\bm{V}}^s)$
        
        $\tilde{\bm{V}}^{(k)}\leftarrow \bm{V}^{(k-1)}+\Delta \bm{V}$\
        
        $\bm{V}^{(k)}=\Omega_{\mathcal{D}}\{\tilde{\bm{V}}^{(k)}\}$
        
        $k = k + 1$
    }


    $\bm{V}_t = \bm{V}^{(K)} $ 
    
    generate $f(\bm{u}) \text{with}\; \bm{V}_t, \bm{w}_t$
    
    $F(\bm{V}_t,\bm{u}_t,\bm{w}_t) \gets \bm{u}_t , \bm{w}_t, \bm{V}_t$
    

    \While{$s\le t/t_{up}$}
    {
    $\mathcal{L}_{F}^s=\frac{1}{t_{up}}\sum_{t_s=(s-1)t_{up}+1}^{st_{up}}\omega_{t_s}F(\bm{V}_{t_s},\bm{w}_{t_s},\bm{u}_{t_s})$
    
    $\bm{\theta}_{\bm{V}}^{s+1}=\bm{\theta}_{\bm{V}}^s-\alpha_{\bm{V}}\nabla_{\bm{\theta}_{\bm{V}}^s}\mathcal{L}_{F}^s$
    
    
    $\bm{\theta}_{\bm{u}}^{s+1}=\bm{\theta}_{\bm{u}}^s-\alpha_{\bm{u}}\nabla_{\bm{\theta}_{\bm{u}}^s}\mathcal{L}_{F}^s$
    
    $\bm{\theta}_{\bm{w}}^{s+1}=\bm{\theta}_{\bm{w}}^s-\alpha_{\bm{w}}\nabla_{\bm{\theta}_{\bm{w}}^s}\mathcal{L}_{F}^s$
    
    $s=s+1$
    }
    }
\caption{\label{alg:meta-WMMSE}General Structure of the  MLBF algorithm for WSR maximization problem. 
}
\end{algorithm}

\section{Simulation Results}
In this section, the performance of the proposed MLBF algorithm is evaluated by extensive simulations. MLBF algorithm is implemented in Python 3.6 with Pytorch 1.6. The WMMSE algorithm is also implemented in Python 3.6 for comparison. We consider the number of transmit antennas $M=4$ and the number of  single-antenna users $N=4$, with user weights $\alpha_i=1$, $\forall i$. We initialize the MLBF and WMMSE with the same initialization of $\bm{V}$ such that $\text{Tr}(\bm{V}\bm{V}^H)\leq P$. 
Both methods are directly evaluated on 1000  channel realizations generated independent and identically distributed (i.i.d.) from a complex standard Gaussian distribution.

For the MLBF algorithm, we set the maximum number of outer loops to $T=500$ when solving each problem and set the maximum number of inner loops to $K=I=J=10$ when updating each variable within the inner loop.  The update interval $t_{up}$ is set to $5$, thus the maximum update times within each sample in the test set is $S=100$. Our meta-learners employ two-layer LSTM networks with 200 hidden units in each layer. The learning rates  of  Adam \cite{kingma2014adam} for updating the LSTM networks are set to $10^{-4}$. The weight of the outer loop step $\omega_{t_s}$ is set to 1 for all $t_s$. We note that our MLBF algorithm is trained during the optimization process and the final obtained results are presented to evaluate the performance. At the beginning of solving each problem, our networks start from scratch. In this way, the MLBF algorithm is implemented as a solution algorithm to the optimization problem at hand, relieved from the time-consuming training process.

Since complex variables are currently not supported by the deep learning tools, we decompose the complex variables  into their real and imaginary parts. Thus we make an alternative reformulation that maps the complex  computations to the equivalent computations in reals. Let the real  matrix $\bm{V}^{'}=[\bm{V}_{re},\bm{V}_{im}]$ denote the variables updated by the meta-learner network. We employ $\bm{M}_{re}$ and $\bm{M}_{im}$ as the masks matrices to recover the real and imaginary parts, respectively; that is, we set $\bm{V}_{re}=\bm{V}^{'}\bm{M}_{re}$ and $\bm{V}_{im}=\bm{V}^{'}\bm{M}_{im}$. Then, it is explicit that $\bm{V}=\bm{V}_{re}+j\bm{V}_{im}$, and the computations in the complex space, such as in (\ref{global loss function}), can be reformulated as real computations with matrices $\bm{V}_{re}$ and $\bm{V}_{im}$. 

{The vanilla WMMSE algorithm is implemented to solve the beamforming problem following the formulation in \cite{Pellaco2020Deepunfolding}, including the stopping criterion that the deviation $\epsilon$ in the WSR between two iterations satisfies $\epsilon\leq10^{-4}$, and employing the bisection search method for computing $\mu$. Differently from \cite{Pellaco2020Deepunfolding}, to ensure the convergence of the WMMSE algorithm, we set the iterative steps $L$ of the WMMSE algorithm to 100 (the maximum number of iterative steps for WMMSE is set to 6 in previous works \cite{Pellaco2020Deepunfolding,Chowdhury2020UnfoldingWMMSE})}. Furthermore, for each sample, we randomly initialize the algorithm for 10 times and choose the best performance for both of the tested algorithms as the obtained result to mitigate the potential impact of poor initializations.

\begin{figure}[htbp]
\centering
\subfigure{
\includegraphics[width=4cm]{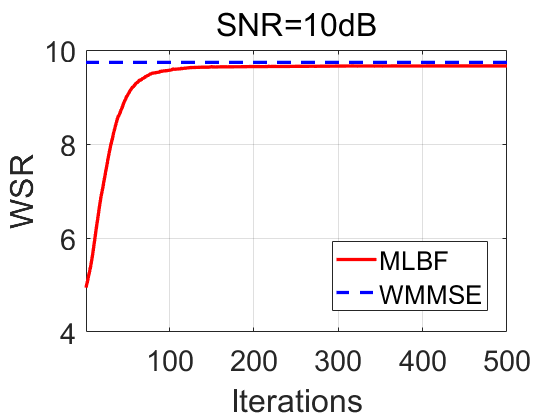}
}
\subfigure{
\includegraphics[width=4cm]{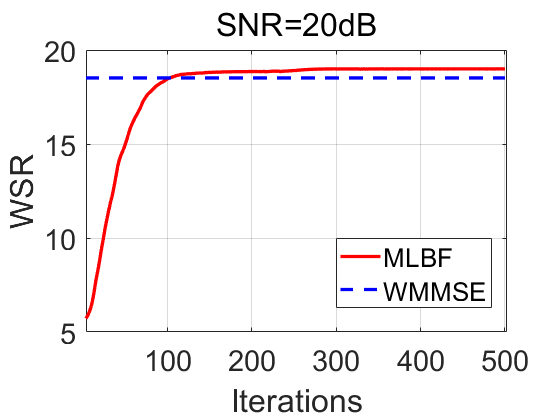}
}
\quad
\subfigure{
\includegraphics[width=4cm]{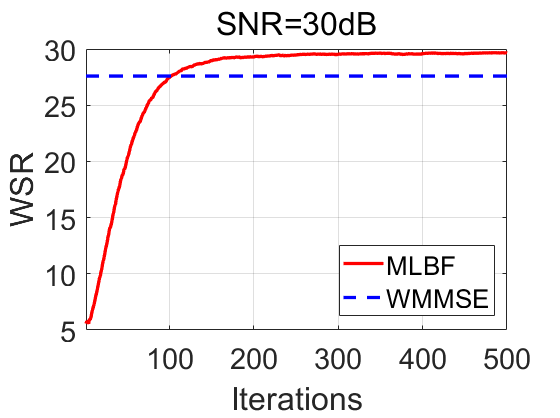}
}
\subfigure{
\includegraphics[width=4cm]{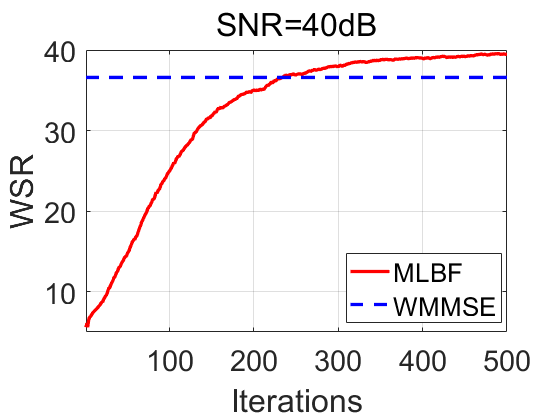}
}
\caption{Red curve denotes the average WSR obtained by the MLBF algorithm, while the blue dash line is the average WSR obtained by the WMMSE algorithm, with L=100. Both  curves are the average WSR results across 1000 channel realizations.}\label{fig:WSR_results}
\end{figure}

\begin{figure}
    \centering
    \includegraphics[width=7.5cm]{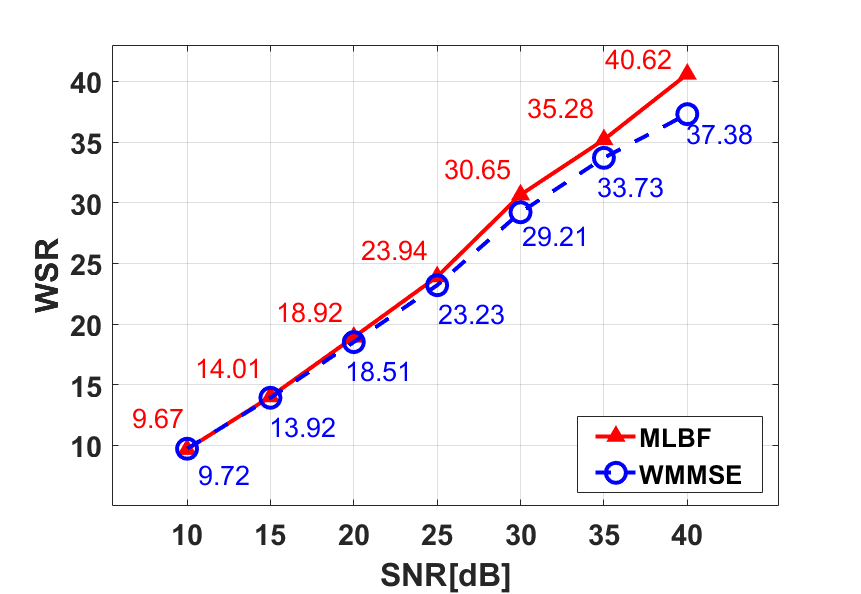}
    \caption{WSR obtained by the MLBF and  WMMSE algorithms as a function of the channel SNR. Each point represents the average WSR across 1000 channel realizations.}
    \label{fig:WSR_compare}
\end{figure}

In Fig. \ref{fig:WSR_results}, we present the obtained  WSR results of the MLBF and WMMSE algorithms averaged over 1000 channel realizations. The red solid line is the WSR results over iterations when using the MLBF algorithm and the dash blue line is from the WMMSE algorithm that is considered as the baseline. It shows that the MLBF algorithm clearly surpasses the WMMSE performance in the high SNR regime of SNR=20-40dB, and achieves comparable performance when SNR=10dB. It can be seen that, at higher SNR, the gap between the WSR obtained by the MLBF algorithm and the WMMSE algorithm becomes larger. In Fig. \ref{fig:WSR_compare} we further show the impact of channel SNR on the obtained WSR by the  MLBF and WMMSE algorithms, respectively. The red solid line with triangles denotes the averaged WSR obtained by the MLBF algorithm as a function of the channel SNR and the blue dash line with circles refers to the results from the WMMSE algorithm. Each point corresponds to the averaged WSR across 1000 channel realizations. It can be seen that when SNR$\leq15$dB, the WSR obtained by the MLBF and WMMSE algorithms are approximately equivalent. With the increase in SNR, the advantage of the MLBF algorithm is more significant. We infer that the low SNR allows both the MLBF and the WMMSE algorithms converge successfully to the optimal solution, thus both algorithms obtain similar WSR results. On the other hand, with the increasing SNR, the non-convexity in the WSR maximization problem is also amplified. Therefore, more local optimal points might exist in the geometry surface of the WSR maximization problem, and the WMMSE is surpassed by the MLBF algorithm significantly as the latter has a more effective search algorithm along the complex geometry.

\section{Conclusion}
In this paper, we proposed a meta-learning based solution to the WSR maximization problem in downlink MISO broadcast channels. The proposed MLBF algorithm is built upon the iterative  block-coordinate descent method, but instead of greedily optimizing for each individual variable at each step, it can adaptively optimize each variable with respect to the geometry of the objective function. In this way, the proposed MLBF algorithm seeks to find a better update of individual variables for the NP-hard and non-convex WSR maximization problem compared to exhaustively minimizing each sub-problem. Simulation results confirm that the MLBF algorithm outperforms WMMSE particularly in the high SNR regime, and achieves a comparable performance when SNR is small. Future works will  focus on  extending the MLBF algorithm to the multiple-input multiple-out (MIMO) and interference channel scenarios, and the consideration of more general utility functions as studied in \cite{shi2011iteratively}. We expect that the gains from the proposed meta-learning approach will be even more significant in those problems due to the increased non-convexity and complexity.
\section{Acknowledgement}
{This work is supported by National Natural Science Foundation of China, projects 61921001 and 62022091, and by the European Research Council (ERC) project BEACON (grant no.677854). We also acknowledge the code from \cite{Pellaco2020Deepunfolding} for the WMMSE algorithm implementation that has been shared by the authors.}
\newpage
\bibliographystyle{IEEEtran}
\addcontentsline{toc}{section}{\refname}\bibliography{Bib_WMMSE}
\end{document}